# A mode-coupling theory analysis of the rotation driven translational motion of aqueous polyatomic ions

Puja Banerjee and Biman Bagchi*

*Solid State and Structural Chemistry Unit, Indian Institute of Science, Bangalore, Karnataka-560012, India*

## *Abstract*

In contrast to simple monatomic alkali and halide ions, complex polyatomic ions like nitrate, acetate, nitrite, chlorate etc. have not been studied in any great detail. Experiments have shown that diffusion of polyatomic ions exhibits many remarkable anomalies, notable among them is the fact that polyatomic ions with similar size show large difference in their diffusivity values. This fact has drawn relatively little interest in scientific discussions. We show here that a mode-coupling theory (MCT) can provide a physically meaningful interpretation of the anomalous diffusivity of polyatomic ions in water, by including the contribution of rotational jumps on translational friction. The two systems discussed here, namely aqueous nitrate ion and aqueous acetate ion, although have similar ionic radii exhibit largely different diffusivity values due to the differences in the rate of their rotational jump motions. We have further verified the mode-coupling theory formalism by comparing it with experimental and simulation results that agrees well with the theoretical prediction.

# I. INTRODUCTION:

Diffusion of ions is of great interest because of their importance in physical and chemical industry, biology and other areas[1-3]. Although many interesting past theories and simulations have focused on structure of solvation shell and dynamics of rigid monatomic alkali cations and halide ions [4-12], a large number of chemically and industrially important electrolytes are made of polyatomic ions such as $NO_3^-$, $SO_4^{2-}$ etc. that has not received much attention. In aqueous solution, these ions with their distributed charges display unusual rotational and translational mechanisms while surrounded by hydrogen bonded network structure of water.

Translational motion of a solute through viscous (Newtonian) fluids is described by Stokes-Einstein relation[13] which is based on Einstein's theory of Brownian diffusion and Stokes' formula for the viscous drag force on a spherical solute in a fluid.

$$D = \frac{k_B T}{C \eta r} \tag{1}$$

where D is the diffusivity of a solute with radius r, $\eta$ is the viscosity of the medium, $k_B$ is the Boltzmann's constant and T is the absolute temperature. C is a numerical constant here that depends on the boundary condition in the particle-fluid interface, being $4\pi$ for slip and $6\pi$ for stick hydrodynamic boundary condition. Though it was originally derived in macroscopic hydrodynamic model, it works quite well for molecular sized particle. But many experimental and theoretical studies have shown that this relation breaks down in supercooled liquids and for spherical monatomic ions in almost all the solvents when the size of the solute is smaller than solvent. This breakdown is more pronounced in the form of Walden's product. According to Walden's rule [1,3], for a particular solute with radius R, the product of limiting ionic conductivity ($\Lambda_0$) and viscosity of the solvent ($\eta_0$) is constant:

$$\Lambda_0 \eta_0 = C/R \qquad (2)$$

where C is a constant depending, among other things, on the hydrodynamic boundary constant, and R is the radius of the diffusing species. Experimental results have shown that for rigid, monatomic alkali ions Walden's product has a maximum for both protic and aprotic solvents when plotted against $R^{-1}$[7]. To describe this anomaly, a number of explanations have been proposed having different theoretical backgrounds. The oldest one is the solvent-berg model that was quite successful. This theory maintains the classical picture of Stokes' law with an *effective* ionic radius assuming the existence of a rigid solvation shell around an ion [1,3]. The other most-discussed model calculates ionic friction by measuring the dielectric response of the solvent when it is perturbed with the electrostatic field of an ion. An ion in a polar solvent induces a polarisation of solvent molecules around it. When the ion moves this polarisation gets perturbed and it undergoes a relaxation process to come to an equilibrated state in the new position of the ion. This dissipated energy during the relaxation process results in an extra friction on the movement of ion that is named as *dielectric friction* by Born [14] many years ago. Hence, the total friction ($\zeta_{total}$) experienced by the ion moving through the viscous medium can be written as:

$$\zeta_{total} = \zeta_{stokes} + \zeta_{DF} \qquad (3)$$

Where $\zeta_{stokes}$ is the friction arising from the Stokes' law due to the shear viscosity of the solvent (friction from short-range interaction) and $\zeta_{DF}$ is the dielectric friction (due to ion-dipole interaction). This model was further developed by Boyd [15], Fuoss [16] and Zwanzig [17], but the final expression proposed by Zwanzig for dielectric friction is found to overestimate the friction of small ions [7]. To rectify this, Hubbard and Onsager (H-O) [18] revised the theory and proposed a set of continuum electrohydrodynamic equations by including both electrostatics and hydrodynamics. The predicted values of ionic conductivity by H-O theory are in a good agreement with experiment up to intermediate ion size but failed to explain the behavior of small ions.

To overcome the limitations of the continuum theories, Wolynes[4] made an attempt towards microscopic theories which was later applied by Colonomos and Wolynes[19] to explain the anomalous behaviour of experimental ionic mobility at zero ionic strength. According to Kirkwood's formula[20], the total dielectric friction on a Brownian ion is given by force-force time correlation function

$$\zeta = \frac{1}{3k_B T} \int_0^\infty dt \langle F(0) \cdot F(t) \rangle \qquad (4)$$

where F(t) is the force exerted on the ion at time t due to ion-dipole interaction. $k_B$ is the Boltzmann constant and T is the absolute temperature. The total force acting on an ion can be decomposed into short range repulsive (hard:H) and long range attractive parts (soft:S) as shown by Allnatt and Rice[21]. Therefore, ζ(t) can be written as a sum of four distinct force-force autocorrelation functions.

$$\langle F(t) \cdot F(0) \rangle = \langle F^H(t) \cdot F^H(0) \rangle + \langle F^S(t) \cdot F^S(0) \rangle + \langle F^H(t) \cdot F^S(0) \rangle + \langle F^S(t) \cdot F^H(0) \rangle$$

$$\zeta = \zeta_{HH} + \zeta_{SS} + \zeta_{HS} + \zeta_{SH} \tag{5}$$

Wolynes identified $\zeta_{HH}$ as Stokes' friction and $\zeta_{SS}$ as dielectric friction and neglected the cross terms. This microscopic theory was further reformulated and generalized by us by taking into account ultrafast solvation dynamics of dipolar liquids using mode-coupling theory formalism[12,22-28].

The significance of the separation of forces into soft and hard parts and their cross-correlations has been discussed in Ref [12]. The cross correlation has been shown to enter into the expression of total force-force correlation function through terms such as $\langle a_{00}(-k) \cdot a_{10}(k,t) \rangle$ which is non-negligible if the soft force contains a radially symmetric attractive terms. The computer simulation study Na+ and Cl- in water by Berkowitz and Wan [29-31] indicated that these cross-correlations is important as the relaxation times characteristic of the soft and hard forces are not widely separated. They computed cross-correlation terms between hard and soft forces for the ion-solvent interactions, the magnitude of which turned out to be non-zero, even significant. Later Tembe and coworkers [32,33] and Koneshan *et al*. [34] also confirmed that cross terms cannot be neglected.

The importance of coupling between the motion of solvent molecule and the solute (ion) was demonstrated by us in Ref. [35]. The generalized theory also treated the role of translational diffusion of solvent molecules on the solute dielectric friction consistently. Translational motion of the solvent molecule can reduce zero frequency dielectric friction (that gives the diffusion) on the solute. In fact, a notable strength of MCT formalism is its ability to treat relaxation of the force on ion by both rotational and translational modes. In addition, MCT can include contributions from both self and collective motions.

Although all these theoretical approaches attempted to explain ionic friction of spherical monatomic ions, no theoretical study was focused to explain dynamics of polyatomic ions in dipolar solvent. In some earlier studies, Barrat and Klein [36] and Lynden-Bell *et al.* [37] analysed translation-rotational coupling of polyatomic ions in solid state ionic crystals. However, too many factors come into the picture if we study polyatomic ions in a dipolar solvent as it shows a complex dynamical motion due to the coupling of solvent motion and solute motion as well as the coupling between translation and rotational motion of polyatomic solutes. This gives rise to many anomalous behaviors leading to a breakdown of Stokes-Einstein relation as diffusivity doesn't vary linearly with the ionic sizes. We can study these systems by no theoretical framework except *mode-coupling theory*.

Polyatomic ions are different from spherical monatomic ions by mainly two factors: 1) their charges are distributed among all constituent atoms and 2) these ions can have certain symmetries, such as $D_{3h}$ ($NO_3^-$), $C_{2v}$ ($NO_2^-$) etc. Therefore, they show a complex dynamics in polar solvents like water in which rotational jump motion of polyatomic ions is coupled to the jump motion of the solvent. Further, the distribution of charges on constituent atoms or groups decides the nature of rotational motion which is again coupled to the translational motion of those ions. Recently, all these properties have been analysed by molecular dynamics simulation [38,39]. *These jump motions introduce a pathway in the system to decrease the friction on an ion which in turn enhances the diffusivity of the ion. This phenomenon can be shown via mode-coupling theory analysis.*

Mode-coupling theory was developed several decades ago, mainly by Götze, Sjögren, Sjölander and their coworkers [40-44]. It was used widely to study glassy dynamics of supercooled liquids. MCT remains the most sophisticated and most successful microscopic theory build from the basic principles of statistical mechanics.

Later, another variant of mode-coupling theory was developed to investigate both polar and non-polar solvation dynamics [45-49] that we will follow to derive MCT formalism for aqueous polyatomic ions in this article. This theory combines density functional theory with time correlation function formalism to derive expressions which when augmented by self-consistency, is essentially an MCT theory. Using this theoretical approach, we accurately explained ion conductance upto a fairly high concentration. In the limit of low concentration this theory reproduced Debye-Huckel-Onsagar expression of conductance, Debye-Falkenhagen expression for frequency dependent conductivity and Onsagar-Fuoss expression for the concentration dependence of viscosity [8,50]. But, all these works aim to study spherical monatomic solutes.

Note that extension of MCT to treat molecules is considerably more difficult because of the presence of orientational correlations. Götze and coworkers studied the motion of a single linear molecule in a liquid of spherical atoms[51,52]. Another form of molecular mode coupling theory (MMCT) was developed by Schilling and coworkers. They studied glass transition of molecular liquid of linear and rigid molecules [53,54] and also applied this theory to water [55]. In this theoretical framework of MMCT they applied tensorial formalism that allows the separation of translation and rotational degrees of freedom. On the other hand, Chong and Hirata [56] applied a site-site description of molecules using the theory of RISM (reference interaction site model) developed by Chandler and Anderson[57]. In case of molecules, different shapes can change degrees of freedom of movement of the molecules. It was demonstrated already that rotational motion can alter the stress for translational motion of center-of-mass of an anisotropically shaped molecule[58-62].

Recently, Schweizer and coworkers studied the coupling of translation and rotational motion of uniaxial hard objects by generalizing naïve mode coupling theory (NMCT) for ideal kinetic arrest and non-linear langevin equation theory (NLE) of activated single particle barrier hopping dynamics to predict translational glass and orientational glass[63]. Using this theoretical approach, they have shown the possibility of multiple glassy states having different ergodicity for rotation and translation: fluid, plastic glass and double glass. They showed that the localization of either CM or rotation or both is determined by solving a coupled equation with translation (localization length, $r_{loc}$) and rotation (localization angle, $\theta_{loc}$).

In case of aqueous polyatomic ions, rotational hopping motion of the ion being coupled with the translation motion of the surrounding water molecules dictates the diffusivity of the ion. In an quantitative description, a self-jump rotation should add a rate of relaxation of the translational memory kernel. This hopping cannot be introduced in hydrodynamic or kinetic theory description but only by mode-coupling theory. In an elegant mode-coupling theory description, we combined translational motion and hopping motion for the activated events in supercooled glassy liquid by using random first order transition theory [64,65].

In this study, we have carried out molecular dynamics simulation of aqueous nitrate and aqueous acetate ions which have similar ionic radius. We found that the rotational motion of nitrate ion (with symmetric charge distribution) is faster than the acetate ion (with asymmetric charge distribution) which in turn enhances translational motion of nitrate ion by a large factor over acetate ion. This agrees well with the experimental findings. In addition, we have modeled some polyatomic ions with the same geometry and non-bonded interactions of nitrate ion but with different charge distribution to further verify the effect of charge distribution of polyatomic ions in their diffusion. This analysis is reported in detail in Ref. [39]. We found that as we break the symmetry of charge distribution of nitrate ion, rotational motion gets hampered and this reduced rotational motion decreases diffusivity of ions.

Motivated by the simulation results, in this article, we discuss similar mode-coupling theory approach to study the anomalous diffusivity of polyatomic ions in water and we compare the theoretical prediction with that of experiment and simulation. We start with mode-coupling theory formalism by including hopping or rotational jump motion into the calculation of ionic friction in section II that can satisfactorily describe the anomalous behavior of polyatomic ions in water. Then we connect this theory to the experimental results and simulation of some polyatomic ions in water in section III and IV respectively. In section V we have shown a semi-quantitative mode-coupling theory analysis of experimental and simulation results and then we conclude our discussion.

## II. MODE-COUPLING THEORY FORMULATION :

In order to motivate subsequent approximations, we discuss mode-coupling theory formalism starting with simpler systems. Mode-coupling theory for a system of a tagged neutral spherical particle in a neutral monatomic solvent system is well established. The exact microscopic expression for the friction on such particle is given by

$$\zeta(z) = \frac{1}{k_B TmV} \int ds\, ds'\, dv\, dv' \left[ \hat{\mathbf{k}} \cdot \nabla_{r_s} u(r_s - r_v) \right] G^s(sv; s'v', z) \left[ \hat{\mathbf{k}} \cdot \nabla_{r_s'} u(r_s' - r_v') \right] \quad (6)$$

where solute is denoted by subscript "s" and solvent is denoted by subscript "v". $u(r_s - r_v)$ is the interaction potential between the solute and the solvent molecules. $G^s(sv; s'v', z)$ is the Laplace transform of Green's function, $G^s(sv; s'v', t')$ which is the four point correlation function that describes the correlated time evolution of the tagged solute (s) and the surrounding solvent molecules (v). It actually measures the probability that the solute moves from $(r_s, p_s)$ at time t to $(r_s', p_s')$ at t′ and the solvent particle located at $(r_v', p_v')$ at t′ and that the same or another solvent particle is found at $(r_v, p_v)$ at time t.

At a microscopic level, the difficulty to study the motion of a solute in liquid arises from the pronounced short range order in spite of having an isotropic structure on the long length scale. Hence, the collisions are strongly correlated. The part of friction that comes from short time direct collision between solute and solvent molecules due to short range interactions (binary collisions) can be approximated by the Enskog binary collision expression [66]. The other contribution comes from a long time process of the correlated recollision of the tagged particle with the same solvent molecules. This arises from the coupling of the solute motion with the different hydrodynamic modes of the solvent. Using the separation of time scales between binary collision and repeated collision (or, ring collision) between solute and solvent molecules, the above expression of friction can be written as

$$\zeta(z) = \zeta_{bin}(z) + \zeta_R(z) \tag{7}$$

Where $\zeta_{bin}(z)$ is the contribution to friction from binary collision. It is the short time part of the friction, the calculation of which is highly nontrivial for a continuous potential. We can approximate it at z=0 by Enskog value of friction ($\zeta_E$) [66]. This calculations are described in detail in the references [41,45].

$\zeta_R(z)$ is the long time part of the friction that arises due to correlated re-collisions of solute particles with solvent molecules and given by the following expression [41,45]:

$$\zeta_R(z) = R_{\rho\rho}(z) - \left[\zeta_{bin}(z) + R_{\rho\rho}(z)\right] R_{TT}(z)\zeta(z) \tag{8}$$

Where $R_{\rho\rho}(z)$ is the coupling of the solute motion to the density modes of the solvent through the two particle direct correlation function, $c_{sv}(k)$. This quantity is obtained though the Laplace transformation of $R_{\rho\rho}(t)$ which is defined as:

$$R_{\rho\rho}(t) = \frac{\rho k_B T}{m} \int \left[ \frac{d\mathbf{k'}}{(2\pi)^3} \left(\hat{\mathbf{k}} \cdot \hat{\mathbf{k}}'\right)^2 k'^2 \times \left[F^s(k',t) - F^0(k',t)\right]\left[c_{sv}(k')\right]^2 F(k',t) \right] \tag{9}$$

Here the input parameters are: 1) the two particle direct correlation function between solute and solvent, $c_{sv}(k)$, 2) dynamic structure factor of the solute, $F^s(k,t)$, 3) the inertial part of the self dynamic structure factor of the solute, $F^0(k,t)$, 4) dynamic structure factor for the solvent, $F(k,t)$. The direct correlation function, $c_{sv}(k)$ couples dynamic structure factors of solute and solvent.

The term, $c_{sv}(k)*F(k,t)$ modifies the dynamics factor of solvent around the solute. $R_{TT}(z)$ in the **Eq. 8** gives the coupling to the transverse current through the transverse vertex function and is defined as:

$$R_{TT}(t) = \frac{1}{\rho} \int \frac{d\mathbf{k}'}{(2\pi)^3} \left[1 - (\hat{\mathbf{k}} \cdot \hat{\mathbf{k}}')^2\right] \left[\gamma_{dsv}^t(k')\right]^2 \omega_{0sv}^{-4} \times \left[F^s(k',t)C_{tt}(k',t) - F_0(k',t)C_{tt0}(k',t)\right] \tag{10}$$

where $C_{tt}(k,t)$ denotes the transverse component of current-current correlation function and $C_{tt0}(k,t)$ denotes the short time part of the same.

The final expression of the total friction becomes

$$\frac{1}{\zeta(z)} = \frac{1}{\zeta_{bin}(z) + R_{\rho\rho}(z)} + R_{TT}(z) \tag{11}$$

By neglecting $R_{TT}(z)$, which is non-negligible only at high density and low temperature, we can write total friction as:

$$\zeta(z) = \zeta_{bin}(z) + \frac{\rho k_B T}{m} \int dt\, e^{-zt} \int \left[\begin{array}{c} \frac{d\hat{\mathbf{k}}'}{(2\pi)^3}(\hat{\mathbf{k}} \cdot \hat{\mathbf{k}}')^2 k'^2 \\ \times\left[F^s(k',t) - F^0(k',t)\right]\left[c_{sv}(k')\right]^2 F(k',t) \end{array}\right] \tag{12}$$

The total friction is determined in a self-consistent calculation where self-consistency comes into picture through the solute dynamic structure factor, $F^s(k,t)$ as it is related to diffusivity of the solute which again depends on the total friction on it. So, **Eq. 12** has $\zeta(z)$ on both sides and thus it has to be solved by a self-consistent calculation.

The system we have discussed upto now, does not have any rotational motion of either solute or solvent. MCT formalism will be modified if rotational motion is included. Here at first we discuss a system of spherical ion in dipolar solvent molecules where solvent molecules have orientational relaxation but the solute does not have any rotational motion. Then we will go to the complex system of polyatomic ions in water where rotational jump of solute molecule is coupled with that of solvent molecules and we will show how rotational jump of polyatomic solute modifies its translational motion. No rigorous MCT formalism exists till now for polyatomic ions in water because of the difficulty that comes into picture due to orientational motion.

### a. Spherical ion in Dipolar Solvent molecules:

Here an extra friction on the solute comes into picture due to the orientational relaxation of dipolar solvent around the ion that is called *dielectric friction*. The exact microscopic expression for the friction on a monatomic spherical ion in a dipolar solvent is given by

$$\zeta(z) = \frac{1}{k_B T m V} \int ds\, ds'\, dv\, dv' \left[\hat{\mathbf{k}} \cdot \nabla_{r_s} u(r_s - r_v, \Omega_v)\right] G^s(sv; s'v', z) \left[\hat{\mathbf{k}} \cdot \nabla_{r_s'} u(r_s' - r_v', \Omega_v')\right]$$

(13)

Here, $ds = dr_s dp_s$ and $dv = dr_v dp_v d\Omega_v$, $\Omega_v$ denotes orientation of solvent molecules and all the other terms have their definitions discussed before. **Figure 1** shows dipolar solvent molecules around a positively charged ion and dielectric relaxation of one of them from $(r_v, \Omega_v)$ to $(r_v', \Omega_v')$ when ion is moving from position $r_s$ to $r_s'$.

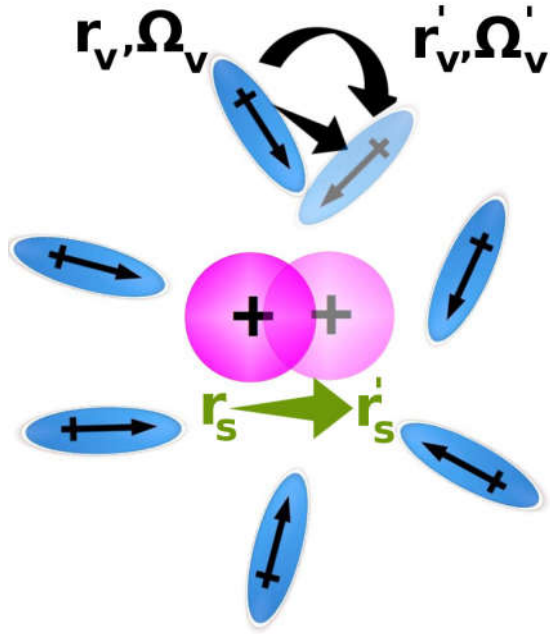

**Figure 1: A Spherical monatomic positively charged ion in dipolar solvent molecules; Orientational relaxation of solvent molecules (subscript: v) in presence of ion (solute; subscript: s).**

The total friction on ions can be written as

$$\zeta_{ion} = \zeta_{stokes} + \zeta_{DF} \tag{14}$$

As discussed before here also, friction arising due to shear viscosity of the solvent that obeys Stokes' law ($\zeta_{Stokes}$) has contributions from two factors, binary collision and ring collision term. Here, this contribution of Stokes' friction is approximated as $4\pi\eta_0 R_{ion}$, where $\eta_0$ is the viscosity of the solvent and $R_{ion}$ is the radius of the ion which is consistent for intermediate sized ions like $Cs^+$ but doesn't obey for small ions like $Li^+$ where size of ion is smaller than the solvent. The second term in **Eq. 14**, $\zeta_{DF}$ is the dielectric friction term that arises due to the long range ion-dipole interaction. It is determined by the ion-solvent force-force correlation function (**Eq. 4**).

By using standard Gaussian approximation, microscopic expression of the dielectric friction is obtained as

$$\zeta_{DF}(z) = \frac{2k_B T \rho_0}{3(2\pi)^2} \int_0^\infty dt e^{-zt} \int_0^\infty dk k^4 F_{ion}(k,t) \left| c_{id}^{10}(k) \right|^2 F_{solvent}^{10}(k,t) \qquad (15)$$

This was first derived in Ref [24]. Here $c_{id}^{10}(k)$ and $F_{solvent}^{10}(k,t)$ are the longitudinal components of the ion-dipole direct correlation function (DCF) and the intermediate scattering function (ISF) of the pure solvent, respectively. $\rho_0$ is the average number density of the solvent. $F_{ion}(k,t)$ denotes the intermediate scattering function of the ion. $F_{solvent}^{10}(k,t)$ is obtained from inverse Laplace transform of $F_{solvent}^{10}(k,z)$ which is expressed as

$$F_{solvent}^{10}(k,z) = \frac{S_{solvent}(k)}{z + \Sigma_{10}(k,z)} \qquad (16)$$

where $\Sigma_{10}$ is the generalized rate of relaxation of dynamic structure factor, given by

$$\Sigma_{10}(k,z) = \frac{2k_B T f(110;k)}{I[z + \Gamma_R(k,z)]} + \frac{k_B T k^2 f(110;k)}{m\sigma^2 [z + \Gamma_T(k,z)]} \qquad (17)$$

where $\Gamma_R(k,z)$ and $\Gamma_T(k,z)$ are the wavenumber and frequency dependent rotational and translational memory kernels, respectively. The term, $f(110;k)$ appeared in **Eq. 17** is defined as orientational correlation function that is related to the direct correlation function, $c(110,k)$.

$$f(110;k) = 1 - \frac{\rho_0}{4\pi} c(110;k) \qquad (18)$$

The contribution of f(110;k) in the single particle limit (k→∞) become unity ( f(110;k)=1), but it has a significant role in the collective density relaxation.

The self-intermediate scattering function of the ion, $F_{ion}(k,t)$ is obtained from inverse Laplace transform of $F_{ion}(k,z)$. In this case, as it is a spherical solute, rate of relaxation of dynamic structure factor ($\Sigma_{10}$) have contribution only from translation motion (**Eq. 16**)

$$F_{ion}(k,z) = \frac{1}{z + D_{ion}^T k^2} \qquad (19)$$

where $D_{ion}^T$ is again determined from the total friction by Stokes-Einstein relation. Therefore, **Eq. 15** being a mode-coupling, nonlinear equation has ζ(z) on both sides that has to be solved self-consistently (as discussed for **Eq. 12**).

### b. Polyatomic ions in water

Polyatomic ions are distinct from rigid monatomic spherical ions by some characteristics key features, most important two among them are distributed charges and orientational caging resulted from hydrogen bonding. Structural fluctuation further manifests itself in the rotational motion of the ions which might include jumps.

Up to this point, the calculation assumed a spherical ion in a dipolar solvent where self consistency comes into the picture though the term $F_{ion}(k,t)$ in **Eq. 15.** In case of polyatomic ions in a dipolar solvent, both the orientational motions of solute and solvent are important to determine the total friction on a polyatomic ion. **Figure 2** shows translational-rotational motion of both the solute and solvent in a system of a polyatomic ion (shown in pink) in water.

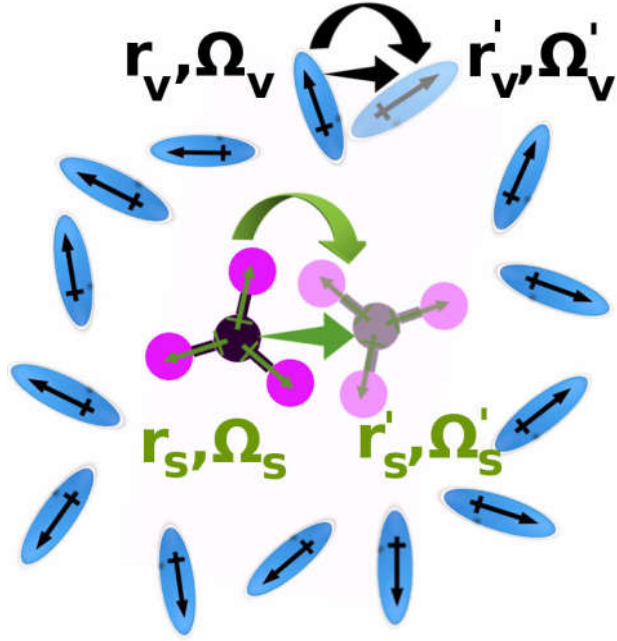

**Figure 2:** A polyatomic ion surrounded by dipolar solvent molecules; Orientational relaxation of both solvent molecules (subscript: v) and solute (subscript: s) while having a translational motion.

In this case, **Eq. 13** is modified as

$$\zeta(z) = \frac{1}{k_B T m V} \int ds\,dv\,ds'\,dv' \left[\hat{\mathbf{k}} \cdot \nabla_{r_s} u(r_s - r_v, \Omega_s, \Omega_v)\right] G^s(sv; s'v', z) \left[\hat{\mathbf{k}} \cdot \nabla_{r_s'} u(r_s' - r_v', \Omega_s', \Omega_v')\right]$$

(20)

Here $ds = dr_s\,dp_s\,d\Omega_s$ where $\Omega_s$ indicates the orientation of polyatomic ions (subscript "s" for solute) and the other terms are defined similarly as discussed before.

A study of mode-coupling theory for polyatomic ions in dipolar solvent is complicated for many reasons. For an aqueous polyatomic ionic system, determination of distribution of solvent molecules around solute molecule is one of the main challenges. One can calculate a microscopic expression of force on the ion using classical density functional theory (DFT) approach where free energy can be written as a functional of position and orientation dependent density of the solvent, $F[\rho(r,\Omega)]$,

$$\beta F[\rho(r,\Omega)] = \int dr d\Omega \rho(r,\Omega)[\ln(\rho(r,\Omega)) - 1] - \frac{1}{2!}\iint dr dr' d\Omega d\Omega' C_2(r,\Omega,r',\Omega')\delta\rho(r,\Omega)\delta\rho(r',\Omega') + \beta U_{ext}$$
(21)

Here $\rho(r,\Omega)$ is the position and orientation dependent density of solvent molecules. The first term on the right-hand side is the contribution of free energy when there is no interaction among the constituents of the system (i.e. ideal gas contribution). Second term comes from two-particle interaction and $C_2(r-r')$ can be identified as the direct correlation function of two sites, $c(r)$ introduced via Ornstein-Zernike equation. Here we have truncated the expression after the quadratic term of density fluctuation. $U_{ext}$ is the total ion-solvent interaction energy. Density functional theory basically provides an expression of the free energy of an inhomogeneous system by a perturbative treatment of the inhomogeneous system in presence of an external potential field, which is given by the presence of the charged solute here.

We can get an expression of force density on the ion using this DFT approach [12,24]:

$$F(r,t) = k_B T\, n_{ion}(r,t)\nabla \int dr' d\Omega' c(r,r',\Omega')\delta\rho(r',\Omega',t)$$
(22)

where $n_{ion}$ is the number density of the ion. We can expand density and direct correlation function ($c(r,r',\Omega')$) in spherical harmonic and using Kirkwood's formula (**Eq. 4**), we obtain frequency dependent dielectric friction as given by the **Eq. 15**.

The theoretical approach discussed above is for spherical solute in dipolar solvent medium. However, in case of polyatomic ions in water, determination of solute-solvent direct correlation function is a nontrivial task. The molecular Ornstein–Zernike (MOZ) theory is a method to calculate three-dimensional (3D) solvation structure in molecular liquids.

$$h(r_1 - r_2, \Omega_1, \Omega_2) = c(r_1 - r_2, \Omega_1, \Omega_2) + \int c(r_1 - r_3, \Omega_1, \Omega_3) \rho(r_3, \Omega_3) h(r_2 - r_3, \Omega_2, \Omega_3) dr_3 d\Omega_3 \tag{23}$$

where pair correlation function of two particles, $h(r_1-r_2,\Omega_1,\Omega_2)$ consists of a direct contribution from direct correlation function between particle "1" and "2", $c(r_1-r_2,\Omega_1,\Omega_2)$ and an indirect part where influence of particle "2" to particle "1" is calculated through other particles. This analytical expression is very hard to solve. Unlike the Ornstein–Zernike(OZ) equation for hard spheres, in molecular liquid it depends on the orientation of two particles also ($\Omega_1$ and $\Omega_2$) which we generally treat through the rotational invariant expansions of interaction potentials and correlation functions. But this formalism does not work well in case of molecular liquids though it works well with the system of prolate-oblate.

Another method is the reference interaction site model (RISM) pioneered by Chandler and Andersen [57] and then it was extended to the polar liquids by the XRISM treatment [67,68]. The theory is based on calculations of radial distribution functions (RDF) via the site–site Ornstein–Zernike (SSOZ) integral equation. According to standard notation of RISM, $\rho_{\alpha M}(r)$ denotes the density at position r of site α for molecules of type M and total pair potential between two molecules, M and M′, u(M, M′) is written as sum of site-site pair-potential denoted as $u_{\alpha M \gamma M'}(r)$ between site α of molecule M and site γ for molecule M′.

$$u(M, M') = \sum_{\alpha=1}^{m} \sum_{\gamma=1}^{m} u_{\alpha M \gamma M'}\left(\left|r_M^\alpha - r_{M'}^\gamma\right|\right) \tag{24}$$

There are two sources of inter-particle correlation in a system of polyatomic ions in dipolar solvent molecules: intra-molecular interaction in both solute and solvent molecules and intermolecular interaction. The RISM integral equation for a general mixture of different species can be written in a matrix form as

$$\rho \mathbf{h} \rho = \mathbf{w} * \mathbf{c} * \mathbf{w} + \mathbf{w} * \mathbf{c} * \rho \mathbf{h} \tag{25}$$

Here $\mathbf{h}$ is the site-site intermolecular pair correlation function matrix, $\mathbf{c}$ is a matrix of direct correlation function and $\mathbf{w}$ is intramolecular correlation matrix. * denotes convolution of different matrices. Matrix $\mathbf{h}$ can be written in terms of different interactions between solute (denoted as "s") and solvent (denoted as "v")

$$\mathbf{h} = \begin{pmatrix} h^{ss} & h^{sv} \\ h^{vs} & h^{vv} \end{pmatrix} \tag{26}$$

The matrices $\mathbf{w}$ has only diagonal terms as per the definition

$$\mathbf{w} = \begin{pmatrix} w^{ss} & 0 \\ 0 & w^{vv} \end{pmatrix} \tag{27}$$

Now, to write an RISM integral equation for a system of polyatomic ions in water we can start from some simpler systems:

a) **Uniform Polyatomic liquid system (water):** Here is only one component in the system, water (denoted as "v") with no solute. Therefore, **Eq. 25** can be written for water-water interaction as:

$$h_{\alpha\gamma}^{\upsilon\upsilon}(r) = w_{\alpha\beta}^{\upsilon\upsilon}(r) * c_{\beta\xi}^{\upsilon\upsilon}(r) * (w_{\xi\gamma}^{\upsilon\upsilon}(r) + \rho_\upsilon h_{\xi\gamma}^{\upsilon\upsilon}(r)) \tag{28}$$

b) **A spherical ion in a dipolar solvent:** There we will have an extra interaction between solute and solvent. But, as the solute is a monatomic spherical ion, site-site intramolecular correlation matrix in the solute particles (denoted by "s") $w_{\alpha\beta}^{ss}(r)$, will be reduced to a scalar quantity (=1)

$$
\begin{aligned}
h_{\alpha\gamma}^{s\upsilon}(r) &= w_{\alpha\beta}^{ss}(r) * c_{\beta\xi}^{s\upsilon}(r) * (w_{\xi\gamma}^{\upsilon\upsilon}(r) + \rho_\upsilon h_{\xi\gamma}^{\upsilon\upsilon}(r)) \\
&= c_{\beta\xi}^{s\upsilon}(r) * (w_{\xi\gamma}^{\upsilon\upsilon}(r) + \rho_\upsilon h_{\xi\gamma}^{\upsilon\upsilon}(r))
\end{aligned}
\qquad (29)
$$

c) **A polyatomic ion in a stockmayer liquid:** If solvent consists of charged monatomic particles, the site-site intramolecular correlation matrix in the solvent particles (denoted by "v"), $w_{\alpha\beta}^{vv}(r)$ will be reduced to a scalar quantity (=1) but there will be a contribution of intramolecular interactions among the sites of the solute (s).

$$
\begin{aligned}
h_{\alpha\gamma}^{s\upsilon}(r) &= w_{\alpha\beta}^{ss}(r) * c_{\beta\xi}^{s\upsilon}(r) * (w_{\alpha\beta}^{\upsilon\upsilon}(r) + \rho_\upsilon h_{\xi\gamma}^{\upsilon\upsilon}(r)) \\
&= w_{\alpha\beta}^{ss}(r) * c_{\beta\xi}^{s\upsilon}(r) * (1 + \rho_\upsilon h_{\xi\gamma}^{\upsilon\upsilon}(r))
\end{aligned}
\qquad (30)
$$

Now, in case of polyatomic ion in water both the solute and the solvent will have intramolecular interactions among their sites. Therefore, the site-site RISM integral equation for the solute-solvent correlations in the system of polyatomic ions in water is

$$
h_{\alpha\gamma}^{s\upsilon}(r) = w_{\alpha\beta}^{ss}(r) * c_{\beta\xi}^{s\upsilon}(r) * (w_{\xi\gamma}^{\upsilon\upsilon}(r) + \rho_\upsilon h_{\xi\gamma}^{\upsilon\upsilon}(r))
\qquad (31)
$$

where "s" and "$\upsilon$" denote solute and solvent particles, respectively as used earlier also. $h_{\alpha\gamma}^{s\upsilon}(r)$ is the pair correlation of solvent site γ around solute site α, $c_{\beta\xi}^{s\upsilon}(r)$ is the site-site direct correlation between solute site β and solvent site ξ, $w_{\xi\gamma}^{\upsilon\upsilon}(r)$ is the intramolecular matrix of solvent $\upsilon$ that depends on the site-separation of $\xi$ and $\gamma$, $l^{\upsilon}_{\xi\gamma}$. $\rho_{\upsilon}$ is the number density of solvent molecules and * signifies convolution. In **Figure 3**, site-site intramolecular and intermolecular interactions have been shown in a pictorial representation in a system of a polyatomic ion with sites α, β and water with sites ξ and γ.

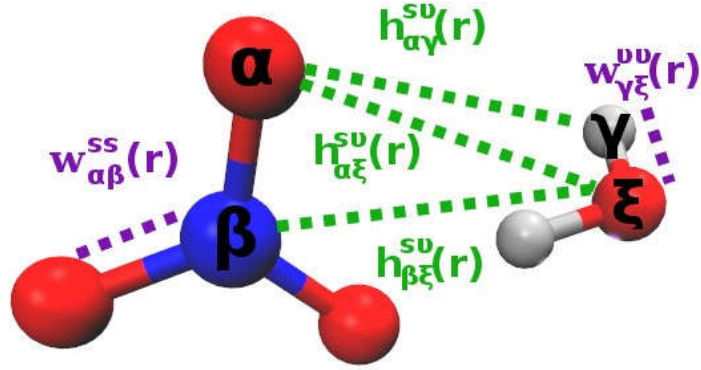

**Figure 3: Interaction between a polyatomic ion (solute) with sites α, β in water (solvent) with sites ξ and γ. Inter-molecular interaction between solute and solvent, $h^{sv}(r)$ and intra-molecular interaction between sites of solute, $w^{ss}(r)$ and solvent $w^{vv}(r)$.**

Chandler, McCoy and Singer [69,70] developed a density functional theory of non-uniform polyatomic system using reference interaction site model (RISM) equation. The density functional truncated to second order in intramolecular and intermolecular correlation is

$$\beta A\left[\langle \rho_{\alpha}(r) \rangle\right]$$
$$= \sum_{\alpha} \int dr \langle \rho_{\alpha}(r) \rangle \ln\left[\frac{\langle \rho_{\alpha}(r) \rangle}{\overline{\rho}}\right] - \Delta\rho_{\alpha}(r) + \beta U_{\alpha}(r)\langle \rho_{\alpha}(r) \rangle - \frac{1}{2}\sum_{\alpha\gamma}\int dr \int dr' \Delta\rho_{\alpha}(r)\overline{c}_{\alpha\gamma}\left(|r-r'|\right)\Delta\rho_{\gamma}(r)$$

(32)

Here, direct correlation function $\overline{c}_{\alpha\gamma}\left(|r-r'|\right)$ is defined in terms of the equilibrium site-site density susceptibility of the uniform unperturbed liquid, $\chi_{\alpha\gamma}(r)$

$$\overline{c}_{\alpha\gamma}(|r-r'|) = (\overline{\rho})^{-1}\delta_{\alpha\gamma}\delta(r) - \chi_{\alpha\gamma}^{-1}(r) \tag{33}$$

The solvent susceptibility is related to density fluctuations of pair of sites $\chi_{\alpha\gamma}(r-r') = \langle\delta\rho_\alpha(r)\delta\rho_\gamma(r')\rangle$. This can also be expressed in terms of intramolecular and intermolecular site-site pair correlation function of the unperturbed liquid. Then by the same approach discussed earlier, we can get the expression for force density on each sites of the polyatomic ion (**Eq. 22**). *Therefore, one can obtain the dielectric friction on the polyatomic ion from DFT-RISM and study mode-coupling theory for friction of polyatomic ions in water.* However, it is nontrivial to do for a real system of a polyatomic ions in water.

In mode-coupling theory, self-motion of ion plays an important role. Similarly, jump rotational motion can help in mitigating the influence of the orientational cage. Therefore, that would be manifested in the self-term of both translational and rotational motion. Now, if we follow the same formalism to that of BBW [64,65] (discussed earlier) to include the contribution of jump motions of polyatomic ions, there will be two rates of relaxation of dynamic structure factor of the solute: First we get from self-consistent mode coupling theory, $K_{MCT}(z)$ as shown in case of spherical ion in dipolar solvent case and another term arises due to the jump rotational motion of the solute, $K_{rot\_jump}(z)$. As they act as two parallel channels for the decay of $F_{ion}(k,z)$, we can write an expression for $F_{ion}(k,z)$

$$F_{ion}(k,z) = \frac{1}{z + K_{MCT}(z) + K_{rot\_jump}(z)} \tag{34}$$

We have already discussed the rate of relaxation of F(k,z) of the dipolar solvent molecules that have contribution from both translational motion and orientational relaxation (**Eq. 16-19**).

According to Hubbard and Wolynes [71], in the limit of slow solute reorientation (when solute rotation is slower than the solvent dynamics), dielectric friction experienced by the solute can be expressed as a time integral of torque-torque time correlation function

$$\varsigma_{DF} = \frac{1}{2k_B T} \int_0^\infty dt \langle \delta\tau(\omega,0) \cdot \delta\tau(\omega,t) \rangle$$

where $\tau(\omega,t)$ is the total torque experienced by the solute molecule with orientation $\omega$ at time t.

In a previous study [35,72] we showed that, using classical density functional theory the expression for dielectric friction on solute due to rotational motion can be obtained as

$$\varsigma_{DF} = \frac{k_B T}{2(2\pi)^4} \int d\mathbf{k} \sum_{m=-1}^{1} \left[ c_{sv}^2(110;k) \langle |a_{10}(k)|^2 \rangle \tau_{10}(k) \right] \qquad (35)$$

with

$$\langle |a_{10}(k)|^2 \rangle = \frac{N}{4\pi}\left(1 + \frac{\rho_0}{4\pi}h(110;k)\right) \qquad (36)$$

where h(110;k) is the Fourier transform of the total pair correlation function in terms of spherical harmonics. The relaxation time $\tau_{10}$ longitudinal (m=0) is given by

$$\tau_{10}^{-1}(k) = 2D_{R0}\left[\left[\frac{k_B T}{2\varsigma(\omega)D_{R0}} + p'(k\sigma)^2\right]\left(1 - \frac{\rho_0}{4\pi}c(110;k)\right)\right] \qquad (37)$$

Here $p' = D_T / 2D_{R0}\sigma^2$, $D_T$ and $D_{R0}$ being translational and rotational diffusivity ($D_{R0} = k_B T/\varsigma_s$, $\varsigma$s is the Stokes' friction) that measure importance of translational diffusion in the polarisation relaxation of dipolar liquid. We found that translational diffusion of solvent molecules can reduce the rotational friction on solute.

But aqueous polyatomic ions, in many cases, exhibit much faster rotational jump motion that cannot be explained by this approach. We are already aware of a formalism which combines mode-coupling theory and hopping. Let us first summarise the formalism given by BBW[64]. The mechanism of hopping is not actually the hopping of atoms/molecule here but the hopping motion from one free energy basin to another free energy basin and the rate of hopping can be expressed as

$$K_{hop} = \frac{1}{\tau_0} exp(-\Delta F / k_B T) \tag{38}$$

where $\Delta F$ is the free energy barrier between two basins that again is related to configurational entropy, $S_c$. This rate, $K_{hop}$ is determined from random first order transition (RFOT) theory.

Now, we write the expression of $F_{ion}(k,z)$ with relaxation time of jump rotational motion ($\tau_R$) of polyatomic ions.

$$F_{ion}(k,z) = \frac{1}{z + D_{ion}^T k^2 + \tau_R^{-1}} \tag{39}$$

Using the expression for $F_{ion}(k,z)$, we can write **Eq. 15** as

$$\zeta_{ion}(z) = \frac{2k_B T \rho_0}{3(2\pi)^2} \int_0^\infty dt e^{-zt} \int_0^\infty dk k^4 \mathcal{L}^{-1}\left[\frac{1}{z + D_{ion}^T k^2 + \tau_R^{-1}}\right] |c_{id}^{10}(k)|^2 F_{solvent}^{10}(k,t) \tag{40}$$

Now, by solving the inverse Laplace transform, this can be written as

$$\zeta_{ion}(z) = \frac{2k_B T \rho_0}{3(2\pi)^2} \int_0^\infty dt\, e^{-zt} \int_0^\infty dk\, k^4 e^{-D_{ion}^T k^2 t} e^{-t/\tau_R} \left|c_{id}^{10}(k)\right|^2 F_{solvent}^{10}(k,t)$$

$$= \frac{2k_B T \rho_0}{3(2\pi)^2} \int_0^\infty dt\, e^{-\left(z+\frac{1}{\tau_R}\right)t} \int_0^\infty dk\, k^4 e^{-D_{ion}^T k^2 t} \left|c_{id}^{10}(k)\right|^2 F_{solvent}^{10}(k,t) \quad (41)$$

Using a different approach, Zhang *et al.* arrived at a similar expression for coupled translational-rotational glassy dynamics of diatomics [63]. But in our approach, we follow a self-consistent calculation to obtain dielectric friction on solute in a dipolar medium. For polyatomic ions in water it is more complicated as we have two species, ions and water and polyatomic ions have charge distributions in different symmetries ($C_{2v}$, $D_{3h}$ etc.). One should have full wavenumber and frequency dependence of self-dynamic structure factor for polyatomic ions in water to develop a mode-coupling theory formalism of such system which is a non-trivial task. Here, we show how rotational jump enhances translational motion of polyatomic ions in water using this formalism. If we can consider two limiting cases:

a) Polyatomic ions can translate randomly, but do not rotate ($\tau_R=\infty$). Therefore, for rotationally frozen (RF) condition diffusivity of the ion is given by zero frequency friction, $\varsigma_{ion}^{RF}(z=0)$ that has the form:

$$\varsigma_{ion}^{RF}(z=0) = \frac{2k_B T \rho_0}{3(2\pi)^2} \int_0^\infty dk\, k^4 F_{ion}(k,t) e^{-D_{ion}^T k^2 t} \left|c_{id}^{10}(k)\right|^2 F_{solvent}^{10}(k,t) \quad (42)$$

b) Polyatomic ions undergo rotational jump motion in water while translating through hydrogen bonded network structure of water. Here the zero frequency friction for the rotationally mobile (RM) ion, $\varsigma_{ion}^{RM}(z=0)$ where $\tau_R \neq \infty$, can be written as (using **Eq. 41** and **42**)

$$\varsigma_{ion}^{RM}(z=0) = \varsigma_{ion}^{RF}\left(z = \frac{1}{\tau_R}\right) \tag{43}$$

which is lower compared to zero frequency friction for rotationally frozen polyatomic ions in water, $\varsigma_{ion}^{RF}(z=0)$ (**Figure 4**). This leads to an enhancement of diffusivity of polyatomic ions due to reduced friction on it, which is caused by rotational jump motion.

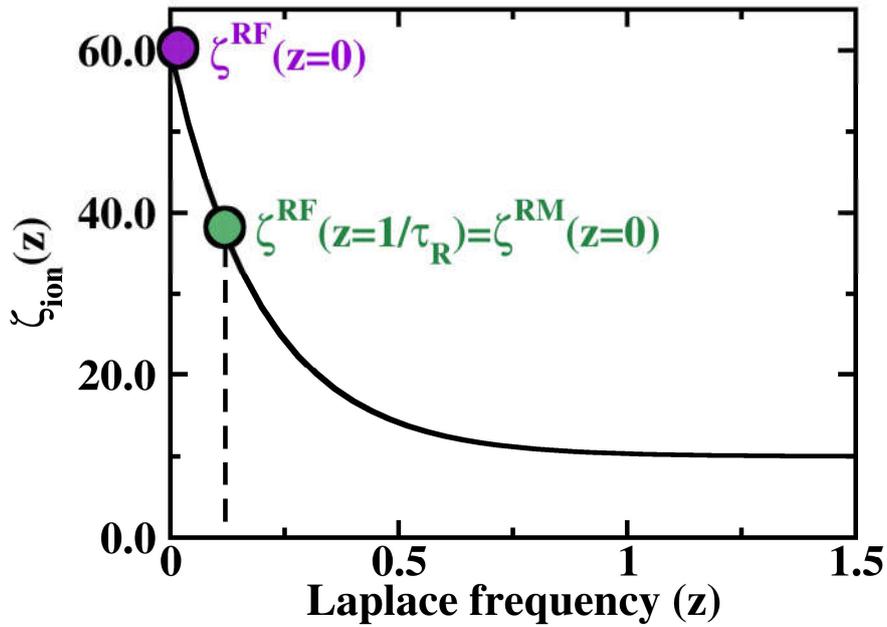

**Figure 4:** Frequency dependent dielectric friction of polyatomic ions in two limiting cases: rotationally frozen(RF) translation motion and rotationally mobile (RM) translational motion.

Therefore, jump motion in turn, contributes to reduce the total dielectric friction on a polyatomic ion that leads to an increase in diffusivity of those polyatomic ions who can easily rotate like nitrate compared to the ions like acetate having reduced rotational motion due to asymmetry in charge distribution. Below we have shown the self-consistent calculation via a flowchart in different types of solute-solvent systems (**Figure 5**).

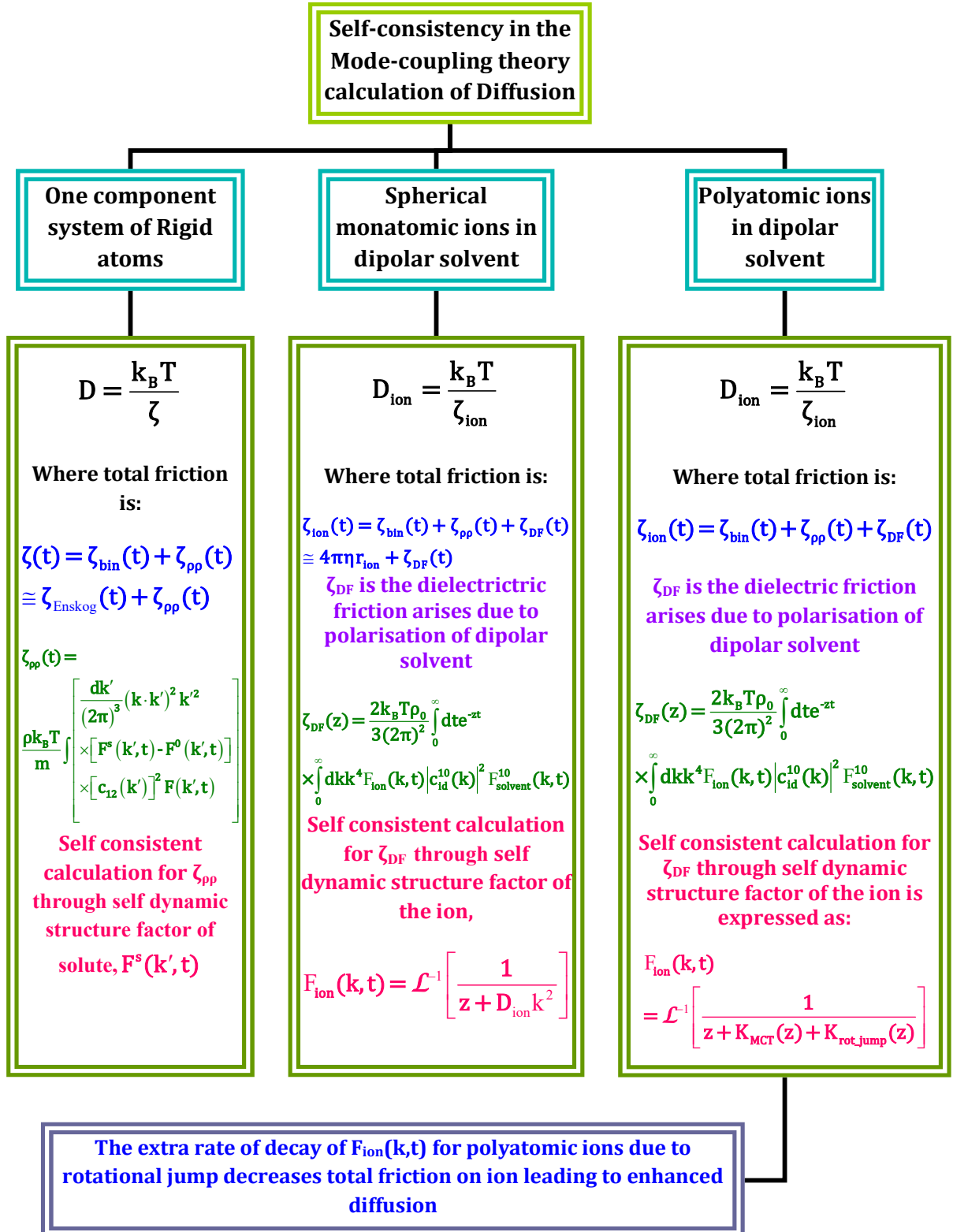

Figure 5: Flowchart showing the difference in self-consistent MCT calculations for different systems.

## III. Analysis of experimental result:

Electrical conductivities of aqueous polyatomic ions have been measured experimentally and the conductivity values at infinite dilution have been reported[73]. Diffusivity values are obtained from experimental conductivity values by Nernst-Einstein relation. We have shown experimental diffusivities of some singly charged polyatomic ions in water at infinite dilution (at $25^0$ C) with their corresponding ionic radius in **Figure 6**. Although these ions have similar ionic radius, the plot shows two separate grouping of ions, one group consists of acetate ($CH_3CO_2^-$), bicarbonate ($HCO_3^-$), iodate ($IO_3^-$) ions with lower diffusivity and the other group consists of nitrate ($NO_3^-$), nitrite ($NO_2^-$), chlorate ($ClO_3^-$), perchlorate ($ClO_4^-$) which have anomalously higher diffusivity values than the ions in the first group.

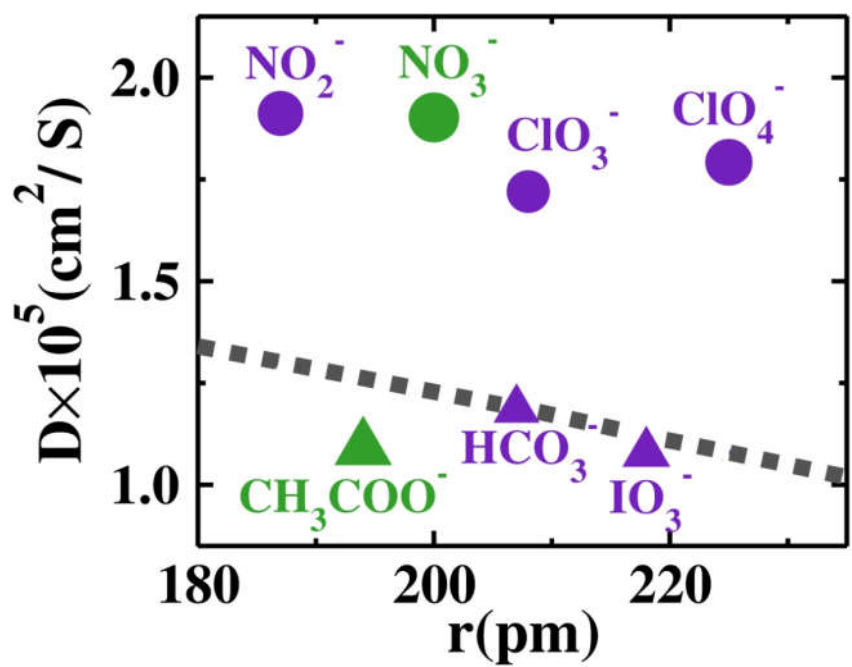

**Figure 6: Diffusivity of chlorate ($ClO_3^-$), nitrate ($NO_3^-$), perchlorate ($ClO_4^-$), nitrite ($NO_2^-$), acetate, bicarbonate ($HCO_3^-$), iodate ($IO_3^-$) at infinite dilution and 25°C (data taken from ref [73]). Diffusivity values are obtained by converting experimental data of conductivity by Nernst-Einstein equation. Dotted line shows the diffusivity**

**values predicted by Stokes-Einstein relation at same temperature and infinite dilution with varying ionic radius of solute.**

Here we can see that an ion like nitrate shows much higher diffusivity from the value of Stokes-Einstein diffusivity where acetate ion does not deviate much from S-E predicted value and also it is not higher than S-E predicted value (**Table I**).

Table I: Comparison of experimental diffusivity values of acetate and nitrate ion with the S-E predicted values

| Species | Experimental diffusivity at infinite dilution and at 298K (cm$^2$/S) | S-E predicted diffusity values at 298K using viscosity of water(0.896 cP) for infinite dilute solution (cm$^2$/S) |
|---|---|---|
| Nitrate ion | 1.902*10$^{-5}$ | 1.22*10$^{-5}$ |
| Acetate ion | 1.089*10$^{-5}$ | 1.25*10$^{-5}$ |

## IV.   Simulation results:

We have carried out atomistic simulations of nitrate ion and acetate ion in water as described in section VII. The reasons behind choosing these two ions are: 1) both have similar ionic radius. Hence, Stokes-Einstein relation predicts similar diffusivity for both the ions, On the other hand, 2) nitrate is a highly symmetric polyatomic ions having symmetric charge distribution over three oxygen atoms where acetate ion doesn't have symmetric charge distribution as well as the structure is also not perfectly symmetric. Therefore, their solvation shells are also structurally different. Symmetric solvation shell gives a symmetric potential energy surface for rotation of the ion while for asymmetric one, potential energy surface becomes distorted. As a result, their dynamics in water is expected to exhibit quite different behavior. Diffusion constants of nitrate ion and acetate ion in water obtained from the simulation at 300 K is reported in Ref. [39]. **Table II** shows the diffusivity values that have a large difference though two ions are similar in size. Nitrate ion exhibits around 78% higher diffusivity value compared to acetate ion which agrees quite well with the experimental results.

Table II: Diffusivity of aqueous nitrate ion and aqueous acetate ion from simulation.

| Species | Size (pm) | Diffusivity of ions in water at 0.1M solution and 300K (cm$^2$/S) |
|---|---|---|
| Nitrate ion | 200 | $(1.67 \pm 0.15) \ast 10^{-5}$ |
| Acetate ion | 194 | $(0.94 \pm 0.06) \ast 10^{-5}$ |

To explain the discrepancy of diffusivity values of these two ions, we have analysed time evolution of the quaternions for a single ion of nitrate and that of acetate at a particular time regions. Both the ions are found to undergo jump rotational motion in water (**Figure 7**). Except $q_4$, all three quaternions show a sudden change together (marked by the black ellipsoids) caused by the rotational jump.

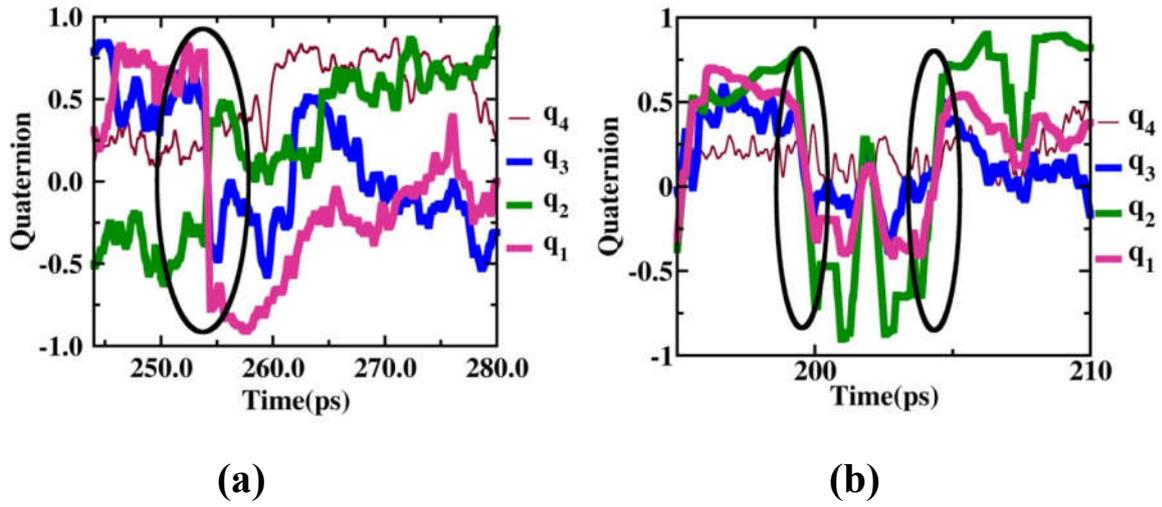

**Figure 7:** (a) Time evolution of the quaternions of that chosen nitrate ion and (b) acetate ion.

Next we investigate the reorientation dynamics of nitrate ion and acetate in water by computing second-order time correlation function, $C_2(t)$ which is defined as:

$$C_2(t) = \left\langle P_2(\vec{n}(t_0) \cdot \vec{n}(t_0 + t)) \right\rangle \tag{44}$$

where $P_2(x)$ is the second order Legendre polynomial defined as

$$P_2(x) = \frac{1}{2}(3x^2 - 1) \tag{45}$$

Here $\vec{n}$ is the unit vector along the bond vectors of nitrate and acetate ions: 3 N-O bond vectors of nitrate ion and 2 C-O and one C-C bond vectors for acetate ion.

In Ref. [39], the time constants for rotational relaxation are reported. For full rotation, decay constant for rotational relaxation of aqueous nitrate ion, $\tau_{NO_3^-}$ = 2.497 ps and that of aqueous acetate ion, $\tau_{Acetate}$ = 6.498 ps. Here, contributions from all three bonds of both the ions have been averaged over to get final $C_2(t)$ values. This result clearly shows that even if these two ions in aqueous medium exhibit a jump rotational motion, the rate of their jump motion is quite different which gives rise to a largely different rate of decay of orientational relaxation of two ions in water.

Now, we measure the probability distribution of mean square displacement of both the ions at two different time intervals(**Figure 8**). The non-gaussian nature of these curves indicates the existance of rotational jump or hopping mechanism involved for both the ions. Also, it is evident that nitrate ion has broader distribution of MSD in both the two timegaps having greater probability to displace a long distance than an acetate ion.

Therefore, there is a qualitative agreement of simulation results with the theoretical formalism, that we discussed in section II. The acetate ion having lower rate of jump motion experiences much more friction compared to nitrate ion, therefore shows less probability of large displacement unlike nitrate ion.

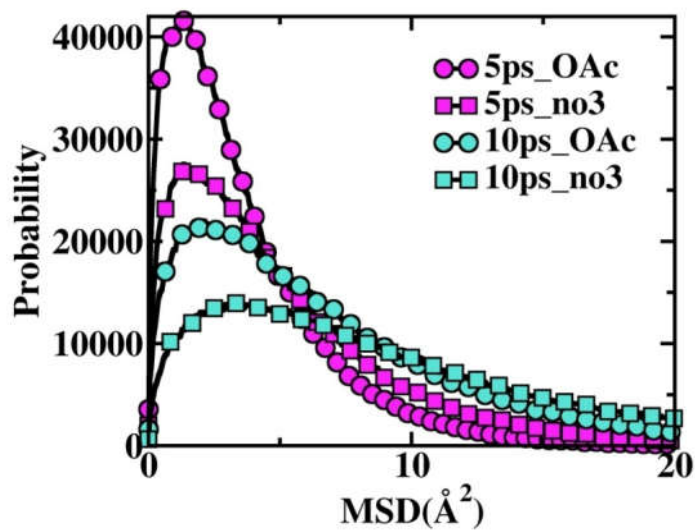

**Figure 8:** Probability distribution of mean square displacements for both nitrate ion and acetate ion.

# V. A semi-quantitative MCT analysis of experimental and simulation results:

As already noted, the full set of MCT equations is exceedingly hard to solve because of the complexity of the orientational pair correlations involved, coupled with the condition of self-consistency. Therefore, a full quantitative analysis of the experimental results is not yet possible. However, we can use the MCT expressions to understand the important role of orientational hopping in augmenting translational diffusion.

From experiments, diffusivities of aqueous nitrate ion and aqueous acetate ion are found to be $D_{NO_3^-}^{expt} = 1.902*10^{-5}$ cm$^2$/S and $D_{Acetate}^{expt} = 1.089*10^{-5}$ cm$^2$/S which have a ratio of 1.75 $\left( = \dfrac{D_{NO_3^-}^{expt}}{D_{Acetate}^{expt}} \right)$ (**Table I**). The same ratio of diffusivities of nitrate ion and acetate ion in water obtained from our simulation is 1.78 $\left( = \dfrac{D_{NO_3^-}^{sim}}{D_{Acetate}^{sim}} \right)$ (**Table II**).

Now, we can take a crude approximation of the **Eq. 41** to obtain zero frequency (z=0) friction on nitrate and acetate ion in water using MCT formalism:

$$\zeta_{NO_3^-}(z=0) = C_{NO_3^-} \int_0^\infty dt\, e^{-\left(\frac{t}{\tau_{R,NO_3^-}}\right)} = C_{NO_3^-} * \tau_{R,NO_3^-} \qquad (46)$$

$$\zeta_{Acetate}(z=0) = C_{Acetate} \int_0^\infty dt\, e^{-\left(\frac{t}{\tau_{R,Acetate}}\right)} = C_{Acetate} * \tau_{R,Acetate} \qquad (47)$$

Where $C_{NO_3^-}$ and $C_{Acetate}$ are two constants for the two ions coming out of the approximation of **Eq. 41**. The constants depend on the diffusivity of these ions, $D_{ion}^T$ in the absence of rotation motion and the direct correlation function between ion and water. Due to the dependence on direct correlation function, the values of $C_{NO_3^-}$ and $C_{Acetate}$ should be different because of different solvation shell structure of these two ions in water, as discussed in Ref. [39].

Now, according to Stokes-Einstein relation, zero frequency friction on a solute is inversely proportional to its diffusivity. Therefore, one can write

$$\dfrac{D_{NO_3^-}^{MCT}}{D_{Acetate}^{MCT}} = \dfrac{\zeta_{Acetate}(z=0)}{\zeta_{NO_3^-}(z=0)} = \dfrac{C_{Acetate} * \tau_{R,Acetate}}{C_{NO_3^-} * \tau_{R,NO_3^-}} \qquad (48)$$

Here $\tau_{R,NO_3^-}$ and $\tau_{R,Acetate}$ are the rotational relaxation time of jump rotational motion of nitrate ion and acetate ion in water respectively which we have measured by computing second order time correlation function, C$_2$(t) of nitrate and acetate ion as discussed in section IV. Using the values of $\tau_R$ obtained from our simulation for these two ions, $\tau_{R,NO_3^-}$ =2.497 ps and $\tau_{R,Acetate}$ =6.498 ps, the ratio from MCT is turned out to be $\frac{D_{NO_3^-}^{MCT}}{D_{Acetate}^{MCT}} = 2.6$, if we assume $C_{NO_3^-} = C_{Acetate}$. This value of the ratio agrees quite well with experimental and simulation result (shown in **Table III**) under a crude approximation. Thus, the MCT formalism explains the enhancement of diffusivity of nitrate ion over acetate ion due to the increased rate of jump rotational motion which is also seen from experiment as well as the simulation.

Table III: Comparion of the value of ratio of diffusivities of nitrate and acetate ions in water obtained from MCT formalism, Experiment and simulation.

|  | MCT predicted value | Experimental result | Simulation result |
| --- | --- | --- | --- |
| $\frac{D_{NO_3^-}}{D_{Acetate}}$ | 2.6 | 1.75 | 1.78 |

We have done a model system study in Ref. [39] by modifying partial atomic charges on the oxygen atoms of nitrate ion. We have found that as the symmetry of charge distribution is broken, it leads to lower diffusivity of the model ions in water. The model system is like nitrate ion but have two different types of oxygen atoms, one Oa and two Ob having different partial charges and we studied 4 different systems with the ratio of partial charges: q$_{Oa}$/q$_{Ob}$= 0.125, 0.5, 1.5 and 2. Below we compare the diffusivity values of those model systems obtained from simulation and from MCT formalism using the rotational relaxation time of those ions obtained from simulation.

Table IV: Comparison of the value of ratio of diffusivities of the model systems with that of nitrate ion ($q_{Oa}/q_{Ob}=1$) in water obtained from MCT formalism and simulation. Diffusivity values and decay constants of rotational relaxation (for full rotation) of the model systems along with that for nitrate ion are shown here (as obtained in Ref. [39]).

| | Diffusivity (cm²/S) | $\tau_R$(ps) | Ratio of Diffusivities from MCT formulation | Ratio of Diffusivities from simulation |
|---|---|---|---|---|
| $q_{Oa}/q_{Ob}=1$ | $1.67*10^{-5}$ | 2.497 | | |
| $q_{Oa}/q_{Ob}=0.125$ | $1.073*10^{-5}$ | 7.802 | $\dfrac{D^{MCT}_{q_{Oa}/q_{Ob}=1}}{D^{MCT}_{q_{Oa}/q_{Ob}=0.125}}=3.12$ | $\dfrac{D^{simu}_{q_{Oa}/q_{Ob}=1}}{D^{simu}_{q_{Oa}/q_{Ob}=0.125}}=1.55$ |
| $q_{Oa}/q_{Ob}=0.5$ | $1.23*10^{-5}$ | 4.394 | $\dfrac{D^{MCT}_{q_{Oa}/q_{Ob}=1}}{D^{MCT}_{q_{Oa}/q_{Ob}=0.5}}=1.76$ | $\dfrac{D^{simu}_{q_{Oa}/q_{Ob}=1}}{D^{simu}_{q_{Oa}/q_{Ob}=0.5}}=1.36$ |
| $q_{Oa}/q_{Ob}=1.5$ | $1.44*10^{-5}$ | 2.936 | $\dfrac{D^{MCT}_{q_{Oa}/q_{Ob}=1}}{D^{MCT}_{q_{Oa}/q_{Ob}=1.5}}=1.18$ | $\dfrac{D^{simu}_{q_{Oa}/q_{Ob}=1}}{D^{simu}_{q_{Oa}/q_{Ob}=1.5}}=1.16$ |
| $q_{Oa}/q_{Ob}=2.0$ | $1.25*10^{-5}$ | 3.745 | $\dfrac{D^{MCT}_{q_{Oa}/q_{Ob}=1}}{D^{MCT}_{q_{Oa}/q_{Ob}=2.0}}=1.5$ | $\dfrac{D^{simu}_{q_{Oa}/q_{Ob}=1}}{D^{simu}_{q_{Oa}/q_{Ob}=2.0}}=1.34$ |

The comparison shows that except $q_{Oa}/q_{Ob}=0.125$, for all the model ions MCT prediction matches reasonably well with the simulation results. We have already mentioned in the discussion of comparison between aqueous nitrate ion and aqueous acetate ion using MCT formalism (earlier this section), that the approximation we have taken here, $C_{q_{Oa}/q_{Ob}=1} = C_{q_{Oa}/q_{Ob}=0.125}$ for **Eq. 48**. It does not obey well for those cases where solvation shell structure is quite different for the two ions. Here, the model system with $q_{Oa}/q_{Ob}$=0.125 resembles to acetate ion according to the partial charges on periphery atoms. Therefore, we see, like acetate ion and nitrate ion comparison, here also for $q_{Oa}/q_{Ob}$=0.125, agreement of MCT formalism with simulation results is not that good like other model ions with $q_{Oa}/q_{Ob}$ =0.5,1.5,2.

Note that, Stokes' friction for polyatomic ions is not well defined although we can consider those ions as rough spheres. But, according to geometry, charge distribution etc. they can behave quite differently. As both simulation and experiment point to the fact that rotational motion of polyatomic ions dictates their diffusivity, we can infer that dielectric friction on those ions which is influenced by the rotational jump of the ion, plays the dominant role to decide the diffusion of a polyatomic ion in water. Here, we have calculated the ratio of diffusivity of two ions by taking the ratio of dielectric friction of the two ions obtained from MCT which gives a good agreement with experimental and simulation results. Therefore, we can conclude that both experimental and simulation results and the trends can be described, semi-quantitatively, by the mode-coupling theory formulation.

## VI. Conclusion

Experimentally, it is found that some polyatomic ions with similar ionic radii exhibit much higher diffusivities than others. Simulation results point to the fact that faster rotational dynamics gives rise to an enhancement of diffusivity of some polyatomic ions and the symmetry of charge distribution plays an important role in this matter. Our main finding is that polyatomic ions in water exhibit rotational jump motions which are coupled to the rotational jump motions of water, and this is responsible for larger diffusivity of some of the ions (not all) depending on the rate of rotational jump motion.

In this article, we have shown that mode-coupling theory can describe this anomalous behavior of polyatomic ions satisfactorily. Hydrodynamics treats rotation and translational motion separately but mode-coupling theory allows us to treat them together and allows us to include the effect of one on the other. Hence, the effect of translation-rotational coupling motion in deciding the diffusion of polyatomic ions can be studied by MCT. The faster structural relaxation after incorporating the rotational jump results in a reduced effective friction on the movement of a polyatomic ion in water that leads to an enhancement of diffusivity. Therefore, according to the rate of rotational jump motion, polyatomic ions exhibit different diffusivity values in spite of having similar ionic radius.

We have taken one ion having high diffusivity value (nitrate), another from low diffusivity branch (acetate) and carried out molecular dynamics simulation in water. Our simulation results reveal that, both ions exhibit orientational jump motion in water and the rate of rotational jump motion ( or, hopping) is higher for a nitrate ion in water than acetate ion that leads to an enhancement of diffusivity of nitrate ion. As a full calculation of mode-coupling theory for polyatomic ions in water is a non-trivial task, as discussed in this article in detail, we have done a semi-quantitative MCT analysis of the experimental and simulation results which shows a good agreement of our MCT formalism with experiment and simulation.

In a previous study, we showed that translational motion of solvent molecules can reduce the rotational friction on the solute, thus enhances its diffusion [35]. Now, in this study, we discussed that rotational motion of polyatomic ions facilitates its translational motion in aqueous medium.

However, rate of rotation of the polyatomic solute in dipolar medium depends on the symmetry of charge distribution that dictates the rate of translation also. The interesting fact to note here is, there is a symmetrical coupling between translation and rotational motion and only mode-coupling theory allows us to study this coupling.

Rotational jump aided translational diffusion has the character of a multidimensional diffusion. This can be understood in the following fashion. When the nitrate ion gets translationally estranged due to an enhanced spatial local order, rotational jumps can provide an escape pathway. It is well-known that diffusion in one dimension can be pathological in the sense that large barrier may not easily be circumvented, leading to low values of self-diffusion [74]. As was shown in Ref. [75], a two dimensional energy surface can greatly enhance diffusion. Actually, the mode-coupling theory expression **EQ. 39** has essentially the same character of a multidimensional diffusion.

Our present study gives rise to some important future problems:

1) Here we have considered only the monovalent ions like nitrate, acetate. But there are many important divalent ions, diffusion of them are important in chemical industries, biological systems etc.

2) Experimentally is has been found that all the halide ions except F- exhibit very high diffusivity which is similar to nitrate, chlorate, nitrite in which case rotation of ion facilitates its translational motion. But the high diffusivity and also the S-E breakdown of halide ions are not understood at present.

## VII.   Simulation details:

Molecular dynamics simulations of potassium nitrate (KNO$_3$) and potassium acetate (CH$_3$COOK) in water have been carried out using the DL_POLY[76] package. Rigid non-polarizable force field parameters have been used for water as well as ions. SPC/E model[77] has been employed for water. For potassium ion in both the simulations, the OPLS-AA[78] force field, for nitrate ion, the potential model suggested by Vchirawongkwin et al have been employed[79] and for acetate ion, we have used united atom OPLS force field [78,80] . The self interaction parameters are listed in **Table V** and consist of Lennard-Jones and Coulombic terms. For both the simulations, 16 cations and 16 anions have been taken in 8756 water molecules in a cubic simulation cell of length 64.04 Å. This corresponds to approximate 0.1 M concentration of aqueous ionic solutions. Simulations were carried out in the microcanonical ensemble with periodic boundary conditions with a cut-off radius of 18 Å. The long-range forces were computed with Ewald summation[81,82]. Trajectory was propagated using a velocity Verlet integrator with a time step of 1 fs. The aqueous KNO$_3$ system was equilibrated for 300 ps at 300 K and then a 2 ns MD trajectory was generated in the microcannonical (NVE) ensemble. The coordinates, velocities and forces were stored every 5 fs for subsequent use for the evaluation of various properties.

Table V: Self-interaction parameters for KNO$_3$ aqueous solution in SPC/E water

| *Species* | *Atom, i* | $\sigma_{ii}$ *(Å)* | $\varepsilon_{ii}$*(kJ/mol)* | $q_i$ *(e)* | *Ref* |
|---|---|---|---|---|---|
| Water | H$^w$ | 0.000000 | 0.000000 | +0.4238 | [77] |
| Water | O$^w$ | 3.166000 | 0.650000 | 0.8476 | [77] |
| Cation | K | 4.934630 | 0.001372 | 1.000 | [78] |
| Nitrate Anion | N | 3.150000 | 0.711300 | 1.118 | [79] |
| Nitrate Anion | O | 2.850000 | 0.836800 | -0.706 | [79] |
| Acetate Anion | $C_1$(CO) | 3.750000 | 0.439000 | 0.700 | [78,80] |
| Acetate Anion | $C_2$(CH3) | 3.910000 | 0.732000 | -0.100 | [78,80] |
| Acetate Anion | O | 2.960000 | 0.879000 | -0.800 | [78,80] |

## ACKNOWLEDGEMENTS

We thank Department of Science and Technology (DST, India), Council of Scientific and Industrial Research (CSIR, India) and Sir J. C. Bose Fellowship to Prof. B. Bagchi for providing partial financial support.